# Ultrafast surface melting of orbital order in $La_{0.5}Sr_{1.5}MnO_4$

Maurizio Monti[1,2*], Khalid M. Siddiqui[1*], Daniel Perez-Salinas[1,3], Naman Agarwal[1], Martin Bremholm[4], Xiang Li (李翔) [5], Dharmalingam Prabhakaran[6], Xin Liu[7], Danylo Babich[7], Mathias Sander[7], Yunpei Deng[7], Henrik T. Lemke[7], Roman Mankowsky[7], Xuerong Liu (柳学榕)[5,8†], Simon E. Wall[1#]

1. *Department of Physics and Astronomy, Aarhus University, Aarhus, Denmark*
2. *Institut des Molécules et Matériaux du Mans, UMR CNRS 6283, Le Mans Université, 72085 Le Mans, France*
3. *ALBA Synchrotron Light Source, Barcelona, Spain*
4. *Department of Chemistry, Aarhus University, Aarhus, Denmark*
5. *School of Physical Science and Technology, ShanghaiTech University, Shanghai 201210, China*
6. *Department of Physics, Clarendon Laboratory, University of Oxford, Oxford, OX1 3PU, UK*
7. *Swiss Light Source, Paul Scherrer Institut, Villigen, Switzerland*
8. *Center for Transformative Science, ShanghaiTech University, Shanghai 201210, China*

*Equal contribution

† liuxr@shanghaitech.edu.cn
# simon.wall@phys.au.dk

## Abstract

Understanding how light modifies long-range order in quantum materials is key to improving our ability to control functionality. However, this is challenging if the response is heterogeneous. Here we address the most common form of light-induced heterogeneity, surface melting, and measure the dynamics of orbital order in the layered manganite, $La_{0.5}Sr_{1.5}MnO_4$. We isolate the surface dynamics from the bulk by measuring the orbital truncation rod as well as orbital Bragg peak. After photoexcitation, the orbital Bragg peak shows an unusual narrowing, which suggests an increase in the correlation length in the probed volume. In contrast, the correlation length at the surface decreases. These differences can be reconciled if the material is heterogeneous, and light melts a less ordered surface. By isolating the surface response, we determine that the loss of long-range order is an incoherent process, which is likely accompanied by the formation of local polarons.

## Main text

Strong coupling between the structure, charge, orbital and spin degrees of freedom results in complex phase diagrams of competing insulating, metallic and magnetic states in quantum materials. As a result, a small change in one degree of freedom can drive a large change in the macroscopic properties of the solid, and using light to initiate non-thermal changes on the ultrafast timescale has emerged as a route to generate new functionalities in these materials[1]. To this end, the ultrafast control of orbital ordering in the manganites has played a key role in understanding and motivating control strategies in a range of quantum materials. Orbital ordering results from a delicate interplay between structural and electron-electron interactions that leads to the manganese 3*d* orbitals forming ordered zig-zag chains which suppress conductivity. However, small changes in the structure or electronic degrees of freedom can drive the system metallic. For example, the manganite $Pr_{0.7}Ca_{0.3}MnO_3$ is an orbitally ordered insulator, but optical excitation can trigger an insulator-metal phase transformation[2,3]. Ultrafast measurements revealed that the Mn-O bond angle was coherently modified by optical excitation, and it was suggested that the change in bond angle favoured the metallic phase[4]. This motivated further experiments to use resonant mid-infrared (mid-IR) light to modulate the crystal structure directly, without exciting the additional carriers produced by optical pulses. This approach was applied to, first, the insulator-metal transition in the manganites[5], and then light-induced superconductivity in cuprates[6], strain engineering in oxide heterostructures[7], and light-induced chirality[8]. In this mechanism, all unit cells are modified in the same way and at the same time through the generation of large-amplitude, long-wavelength distortions[9]. Theoretical calculations showed that such coherent structures could result in new emergent electronic properties[10], demonstrating that material properties can be changed with optically driven structural changes.

To verify the structural changes directly, time-resolved Bragg scattering experiments found changes consistent with the coherent mechanism expected for both optical[11] and mid-IR excitation[12] in the manganites, and the coherent model has, in general, been applied to understand diffraction data in a wide range of different materials[13–16]. However, translating changes in Bragg intensities to actual distortions of the crystal structure is challenging and requires assumptions when only a limited number of Bragg peaks are measured. In many cases, it is assumed that the changes occur due to coherent distortions, the system is homogeneous in-plane, the depth dependence is linear with the pump absorption and the pump and probe depth profiles are known[11–18]. However, recent observations of ultrafast disorder transitions[19–23], where Bragg intensities are suppressed because of the generation of a broad range of incoherent structural distortions, as opposed to coherent changes, question these assumptions.

Disordered short-range structural correlations are increasingly recognised as having a major impact on material functionality[24]. Disordered phases have a high symmetry on average, but locally the symmetry is broken. This can enable macroscopic properties that are not possible if the structure is both globally and locally symmetric[24–26]. These states cannot be accessed through a coherent modification of the crystal structure as this changes both the local and global symmetries. Thus, understanding if a light-induced phase transition is coherent, or disorder-driven, is critical to understanding the origins of light-induced properties. This is particularly important in the manganites because the thermally driven orbital-ordering phase transition is order-disorder-like with local orbital fluctuations persisting above $T_c$[27]. This contrasts with the suggested light-induced coherent mechanism in which the local order is fully suppressed[11]. While this might point towards a non-thermal nature of the light-induced transition, recent all-optical experiments have shown that disorder generation is consistent with the dynamics of the ultrafast transition, if the system is assumed to be heterogeneous[22]. Indeed, such heterogeneity should be common in ultrafast light matter control as the light stimulus is first absorbed in the surface region, naturally introducing heterogeneity. Therefore, we revisit the melting of orbital order in the prototypical manganite, $La_{0.5}Sr_{1.5}MnO_4$ following photoexcitation to measure possible light-induced disorder and heterogeneity.

Coherent and disorder transitions can be distinguished through the weak diffuse scattering that appears between the Bragg peaks[28]. However, interpreting dynamics in heterogeneous samples remains a challenge. Advances in ultrafast X-ray[29] and near-field[30] imaging now enable measurements of in-plane heterogeneity down to the nanoscale on the ultrafast timescale and time-resolved LEED can measure changes in the surface structure[31], but techniques that can separate the surface response from the bulk are lacking. This is particularly needed for X-ray measurements as the depth dependent signal is needed to interpret the bulk response of the material reliably.

To overcome this limitation, we exploit ultrafast X-ray surface and diffuse scattering[32] to access the dynamics that occur exclusively at the crystal surface of a bulk crystal and compare our results to those obtained from a Bragg peak, which is usually assumed to measure the bulk order, within the same experiment. We find that the surface dynamics are in stark contrast to those obtained from the Bragg response and conclude that melting orbital order proceeds via disorder generation and local fluctuations, rather than the previous cooperative pathway involving coherent motions. Thus, this work not only brings about a technical advance needed to capture the dynamics of light-induced processes in a range of materials, but it also exposes a new mechanism by which ultrafast phase transitions can proceed and be controlled.

**Figure 1a** shows the experimental setup used to achieve surface sensitivity. Experiments were performed at the SwissFEL instrument Bernina[33]. The sample was cooled to 180 K, to be in the orbitally ordered phase which has a critical temperature of approximately $T_c$ ~ 220 K. Surface sensitivity is achieved by measuring the (-0.25, 2.25, L) orbital truncation rod (OTR) in the vicinity of L=0, which is the scattering that arises from the orbital-ordered surface. Scattering from this region is maximised in a grazing incidence geometry, which also helps to match pump and probe penetration depths, which were calculated to be roughly equal. To probe the bulk order, we measure the (-0.25, 2.25, 2) orbital Bragg peak (OBP). Both measurements were performed at the same X-ray grazing incidence angle enabling a direct comparison (see Methods for details).

We start by discussing the response of the OBP. Static characterisation through the analysis of the peak width before excitation gave an in-plane correlation length of $\zeta_{ab} = 250$ Å, and out-of-plane correlation of $\zeta_c = 81$ Å, in good agreement with previous measurements[34,35]. **Figure 1b** shows the temporal evolution of a cut through the OBP along the L direction. Upon excitation with 800 nm pump, the peak intensity reduces as the amplitude of the orbital order is rapidly suppressed. However, the width of the remaining peak narrows on the same timescale (**see extended Figure 1**). The narrowing also occurs in the in-plane direction as found from a full three-dimensional scan of the pumped OBP shown in **Fig 1c**.

We note that we could only suppress the OBP intensity by ~30% with a pump fluence as high as 12 mJ/cm$^2$. Recent optical measurements suggest that the order should be completely melted at these fluences[22]. Such discrepancy results from the fact that the X-rays probe some unpumped material. To extract the changes in the pumped region, knowledge of the depth-dependent order is required. As this is typically not measured, the common assumption is that the order was homogeneous before excitation, i.e. the correlation length is independent of depth from the surface[11,17,18]. In this context, the observed photo-induced narrowing of 10% would imply that the residual orbital order must *increase* its correlation length. Such a photo-induced annealing process, while exotic, has been proposed to explain peak width changes in materials with competing charge density wave order[36]. However, if the material starts in a heterogeneous state, alternative interpretations need to be considered.

As the initial heterogeneity is unknown, we do not attempt to model the depth profile of the excitation. Instead, we turn to the OTR, which is only sensitive to the surface region. **Figure 2a** shows cuts through the OTR in equilibrium, which reveal an in-plane correlation length, $\zeta_s$ ~ 40 Å (**see Methods**), consistent with previous measurements[34,35]. The in-plane correlation length at the surface is a factor of ~6 shorter than at the bulk, demonstrating that the correlation length must vary as a function of depth from the surface into the bulk. **Figure 2b** highlights the surface sensitivity of our experiment by comparing the fluence dependence of the raw detected intensities of the OTR and OBP. The surface scattering signal saturates at much lower fluences than the bulk, and the observed ~80% decrease above 4 mJ/cm$^2$ excitation indicates the near-complete melting of orbital ordering at the surface layer, with no further significant changes when the fluence is increased by more than a factor of 2 from 5 to 12 mJ/cm$^2$. Excitation at 16 mJ/cm$^2$ resulted in a persistent change in the initial order at the surface, but also no further suppression of the photoexcited state was observed (see **extended Figure 2**). **Figures 2c-f** show line cuts through the OTR at L=0.2 at several fluences and delays. Here, an increase in diffuse scattering around the OTR can be seen to accompany its suppression, pointing to a disorder driven change[19].

In **Fig. 3a,** we plot the dynamics of the measured peak intensity as a function of time. Like the OBP, we observe a rapid suppression, but now clear oscillations modulate the response. At low excitation fluence, a single frequency response at $\omega = 2.8$ THz is observed. As the fluence is increased the amplitude of this mode is suppressed and a new frequency emerges at the second harmonic, ~5.6 THz (**Fig. 3b**). While coherent oscillations on superlattice peaks are usually taken as an indicator for coherent oscillations of the order parameter, the low frequency mode observed here can be identified as a Raman active phonon due to the motion of the La/Sr ions[37] , which is a spectator mode[11,22] of the phase transition. Therefore, there are two contributing factors to the scattered intensity at the orbital ordering wavevector, one coming from the scattering of the orbital order itself and a second from a phonon response at the same wavevector. The detected intensity is then sensitive to the square of the sum of the scattered fields from these two terms, i.e. $I_D \propto \left(f_{oo} + f_{ph}\right)^2$, where $I_D$ is the detected intensity, $f_{oo}$ is the structure factor for the orbital ordering and $f_{ph}$ is the structure factor of the phonon distortion, which is proportional to the phonon amplitude. In equilibrium, $f_{ph}$ is zero so that the intensity is purely due to the presence of orbital order. Out-of-equilibrium, the value of $f_{ph}$ oscillates and, if $f_{oo}$ remains finite, there will be a cross term that is linear in the phonon amplitude and order parameter. However, if $f_{oo}$ is fully supressed, the detected intensity will be proportional to the square of the phonon amplitude, resulting in the appearance of the second harmonic. A simple model for this process is discussed in the **supplementary information**, which shows that the fundamental mode frequency exhibits a weak softening under intense photoexcitation (**extended data Figure 3)**.

To extract the dynamics of the surface correlation length, we first subtract the coherent oscillation from the data and fit the peak width. **Figure 3c** shows the resulting dynamics of the change in width measured at L=0.2. In contrast to the OBP dynamics, we now find that the width of the OTR increases, meaning that the surface correlation length decreases as a function of time. Interestingly, we find that the magnitude of the width change appears independent of fluence at early times, which is not observed in the amplitude dynamics. We fit the change in correlation length, extracted from the width dynamics, with a phenomenological model that assumes that photoexcitation, in the long-time limit, establishes a new correlation length, $\zeta_p$ and that the rate of change of the correlation length depends on the square of the difference between the current and the final value.

$$\frac{d\zeta}{dt} = -k\left(\zeta - \zeta_p\right)^2 ,$$

where $k$ is the evolution rate constant. To fit the fluence-independent initial dynamics, the evolution rate must decrease for increasing fluence. This slowing is not observed in the dynamics of the peak (see **extended data Figure 4** for a direct comparison) which indicates that changes in the 'amplitude' or structure factor of the orbital ordering distortions are distinct from the changes in the correlation length and that a range of processes on different length and time scales are occurring.

From these observations, we can now reconcile the observed narrowing of the OBP. The OBP measures the average Bragg intensity over the probed depth of ~100 nm, thus it includes a contribution from the surface region, as well as deeper into the bulk. The correlation length at the surface, as determined by the static OTR is about a factor of 6 shorter than in the bulk, thus the initial state of the sample is heterogeneous. The laser strongly supresses the amplitude and the correlation length of the OO near surface region, but the bulk is much less affected. As a result, the surface contribution is driven into the weak diffuse background, and the remaining scattering comes from order deeper in the bulk. Since this region has a longer correlation length before excitation, the OBP narrows, even though no increase in correlation length occurs.

Having isolated this heterogeneous response, we can now understand the true nature of the photo-induced phase transition. To understand the mechanism in more detail, we look at a larger region of reciprocal space around the OTR. **Figures 4a**,**b** show the region around the OTR 1 ps after pumping at 12 mJ/cm$^2$. In addition to the suppression of the OTR (blue), there is a large increase in diffuse scattering spanning a much larger region of reciprocal space (red). The distribution is anisotropic and peaked along the line *[H, K] = [2-K, K],* and is extended in L. The intensity appears to increase towards the (020) structural peak. This is consistent with diffuse scattering from local nanometre scale lattice strain[38] and indicates that a broad distribution of phonons in momentum space is generated, with a maximum at the $\Gamma$-point. The dynamics of diffuse scattering at various points in momentum space, are shown in **Fig. 3e**, illustrating that the rise time depends strongly on momentum. These dynamics can be fitted with an exponential rise, which is shown in **Fig. 4c,d** for different momenta. The dynamics are fastest closest to the (020) structural peak, and become slower when moving away along the *[H, K] = [2-K, K]* diagonal or in L. Although we do not measure closer to the (020) peak, we find that the fastest increase in diffuse signal (115 fs) is already comparable to the ~50 fs, resolution-limited, decay time of the orbital order (dashed line in **Fig. 4d**). The similarities in timescales suggest that the loss of long-range orbital order and rapid generation of incoherent phonons are driven through the same mechanism.

Such rapid changes in intensity near the $\Gamma$-point is surprising. Photoexcited electrons typically relax through scattering with high energy acoustic modes at the Brillouin zone boundary first, and then these modes decay into phonons closer to the $\Gamma$-point by slower phonon-phonon scattering events as seen in charge density wave compounds[31,39]. The observation of high-q scattering before low q-scattering was also attributed to the formation of localised small polarons that grow as a function of time in the lead halide perovskites[40]. This is the opposite to what we observe. Optical measurements show that photoexcitation couples to a broad range of zone centre optical phonons with frequencies up to 20 THz[22]. These modes are initially delocalised, but we conjecture that they rapidly localise around the photoexcited charges. The slower broadening of the diffuse distribution in momentum space is consistent with a localisation of these long-range distortions into disordered local polarons. A similar process has been proposed in SnSe[41], but polaron localisation in the manganites is an order of magnitude faster, most likely due to the involvement of the light oxygen atoms and the strong polaronic coupling in this material.

The picture that emerges for melting of orbital ordering is as follows: before optical excitation, orbital order at the surface is a patchy network of zig-zag orbital chains, with a correlation length of ~60 Å. Optical excitation rapidly and uniformly weakens the amplitude of the distortions that give rise to orbital ordering, without changing the correlation length. The rate equation-like growth of the width suggests the weakened order is first lost at the edges as the domains shrink. During this 'melting' process, polaron localisation occurs. The anisotropy of the diffuse scattering shows that the polaronic distortions are not random but are correlated along the zig-zag orbital-ordering chain direction than perpendicular to it. This suggests that there is still a degree of local orbital order within a chain, but chains are no longer aligned, consistent with a disordered phase transition[27]. We point out that, while the dynamics we observe are consistent with a polaron localisation, the diffuse scattering signal could also arise from delocalised phonons with random phases. True verification of the polaronic nature of this scattering would require knowledge of the phase of the scattered X-rays which we do not get from an intensity measurement. However, this may be possible in the future with coherent scattering techniques[42].

This order-disorder mechanism is in stark contrast to the coherent mechanism as previously proposed[4]. Instead of a uniform change in the crystal structure as predicted by the coherent model, the transition is highly heterogeneous and shares many similarities to the dynamics observed in $VO_2$[19]. In particular, the rapid increase in diffuse scattering immediately after excitation may suggesting that control strategies that exploit transient disorder may also be applicable for optimising transitions in the manganites[21]. Understanding the transient state generated, either polaronic or incoherent phonons origin, and how it can be controlled, will require new theoretical approaches that can handle disorder dynamically[43–45] as well as density functional theory that can accurately capture random local symmetry breaking due to disorder[25,26].

Although the phase transition itself is incoherent, the lattice dynamics of the spectator modes can remain coherent throughout the melting process. This implies that the disappearance of vibrational modes in optical experiments does not necessarily mean all vibrational coherence is lost[23], but rather becomes undetectable as the mode is no-longer optically active. An interesting question remains as to whether resonant IR excitation, which has also been shown to melt orbital ordering[46,47], still takes the coherent pathway or if it too could be driven by disorder. A coherent mechanism would not directly induce the diffuse scattering seen here and thus repeating this measurement under mid-IR excitation could reveal how direct lattice excitation is different from the optical excitation.

We stress that the ability to directly resolve depth-dependent heterogeneity has been key to making these observations. Specifically, interpreting the Bragg peak data at face value gives an incorrect picture of the physical processes occurring. In the case discussed here, the sample started in a heterogenous state, but the pump profile will always result in a heterogeneously excited volume even in samples that are initially homogeneous. Ultrafast melting has been reported without peak broadening which was attributed to the absence of topological defects[48]. Our results indicate the broadening may occur at the melted surface but is hidden from Bragg probes even when the penetration depths of the laser and X-rays are nominally matched. Similarly, the melting of charge density wave order has received significant attention, and the dynamics have commonly been interpreted assuming a homogenously ordered initial state [14–16]. However, recent photoemission measurements show that amplitude mode is different at the surface than in the bulk[49] indicating the possibility that some physics may have been missed in previous reports. Thus, methods to resolve light-induced heterogeneity will be vital to accurately interpret dynamics in quantum materials.

## Acknowledgements

We acknowledge the Paul Scherrer Institut, Villigen, Switzerland, for provision of beamtime at the Bernina beamline of SwissFEL. We also thank Jakob Voldum Ahlburg, the Center for Integrated Materials Research (iMAT) at Aarhus University, and Huang Shih-Wen for support on sample characterisation. We thank the Danish Agency for Science, Technology, and Innovation for funding the instrument center DanScatt for travel support.

Carlsbergfondet (CF20-0169), Novonordisk foundation (NNF23OC0084990 - Harnessing Dynamic Disorder for Efficient Material control (HD-control)). S.E.W
MOST of China (2022YFA1603900). X. Li and Xuerong Liu

## Author Contributions

S.E.W and Xuerong Liu conceived of the project.
Samples were grown and characterized by D.P.
K.M.S and N.A prepared samples for experiments.
K.M.S, M.M, D.P.S, N.A, M.B, S.E.W performed the experiment with support from Xin Liu, D.B, M.S, Y.D, H.T.L and, R. M. Data analysis was performed by M.M, X.Li, Xuerong Liu and K.M.S.

## Competing interests

The authors declare no competing interests.

## Figures

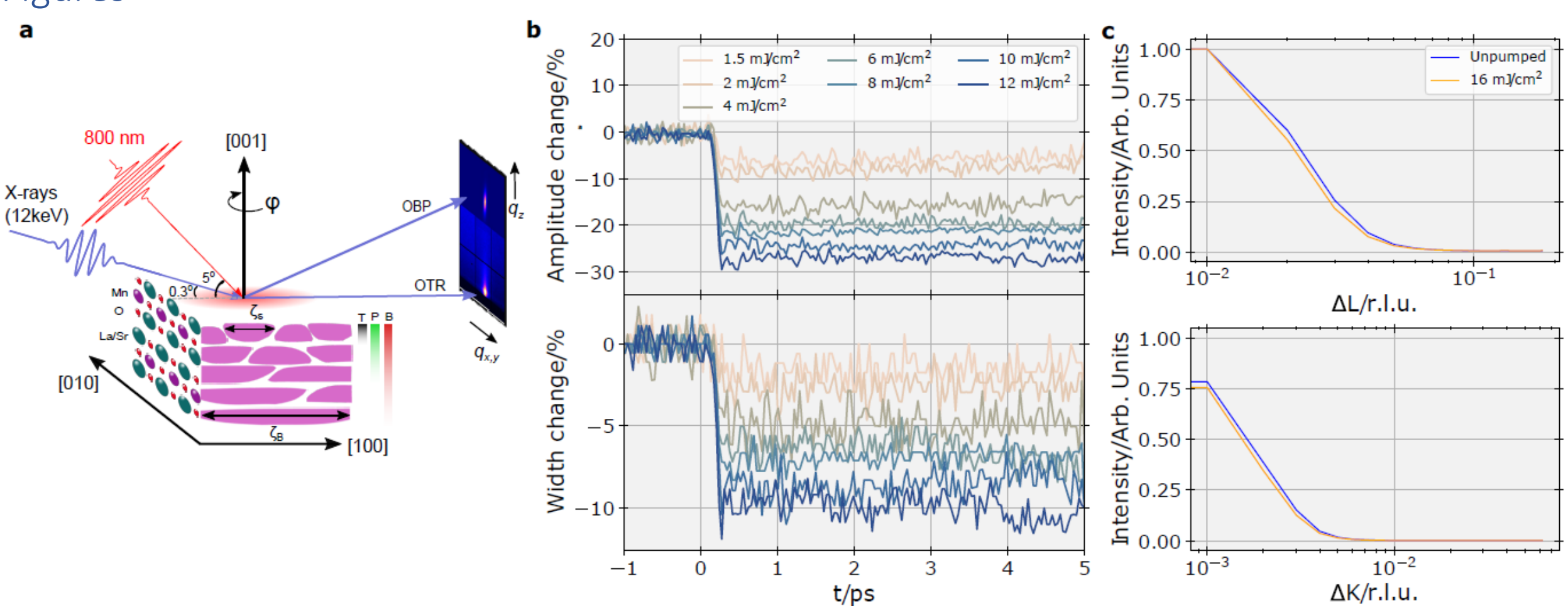

**Figure 1 Experimental setup and dynamics of bulk orbital order. a** Schematic of the experimental setup to measure the (-2.25, 0.25, 2) orbital Bragg peak and (-2.25, 0.25, L) orbital truncation rod. Measurements were performed at 0.3° grazing incidence and switching between the OBP and OTR involved changing the angle $\phi$. $\zeta_s$, $\zeta_B$ indicate the difference in correlation length at the surface and bulk respectively. T, B, P indicate the depth probed by the truncation rod (T) and Bragg Peak (B), and the depth pumped by the laser (P). **b** Dynamics of the amplitude and width along the L direction after photoexcitation. **c** A full reciprocal space map of the OBP at 1 ps delay and 16 mJ/cm² excitation obtained by scanning the $\phi$ angle and maintaining the X-ray angle of incidence. Narrowing is seen along both K and L directions.

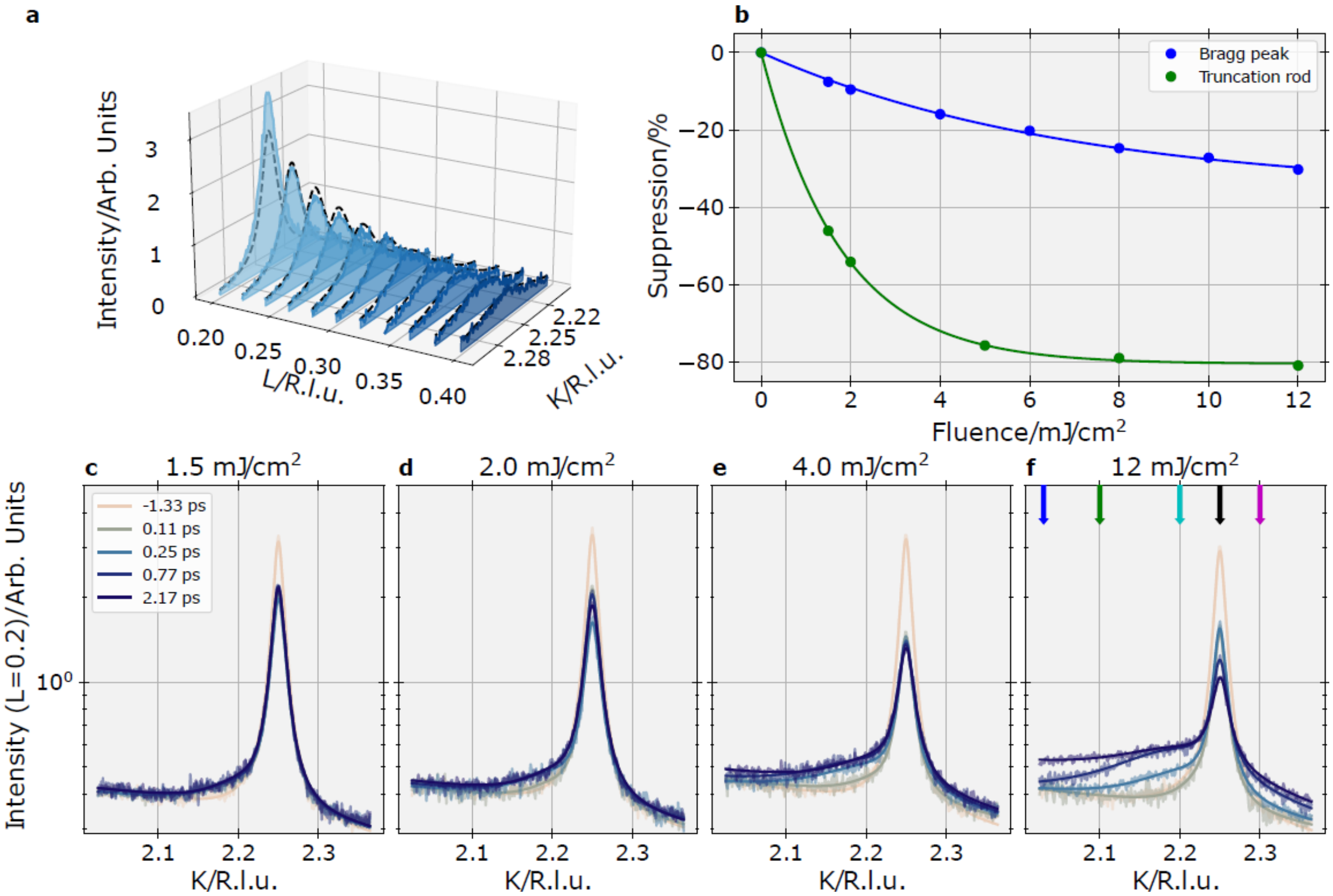

**Figure 2 Orbital ordering melting dynamics at the surface measured through the orbital truncation rod (OTR). a** Measurement of the OTR in the KL plane. L is parallel to the surface normal. Dashed lines are fits based on modelling the surface scattering (see **Methods**). **b** Fluence dependence of the OBP and OTR obtained from integrating a small ROI in reciprocal space around the peak scattering in each case, 1 ps after excitation, solid lines are guides to the eye. **c-f** Line cuts through the orbital truncation rod at L = 0.2 showing that at high levels of excitation, the scattering from the OTR is suppressed and a broad distribution scattering is generated. Solid lines represent a multi-Lorentzian fit to the data (see **Methods**). Time traces at the K points indicated by the coloured markers are shown in **Fig. 3e**.

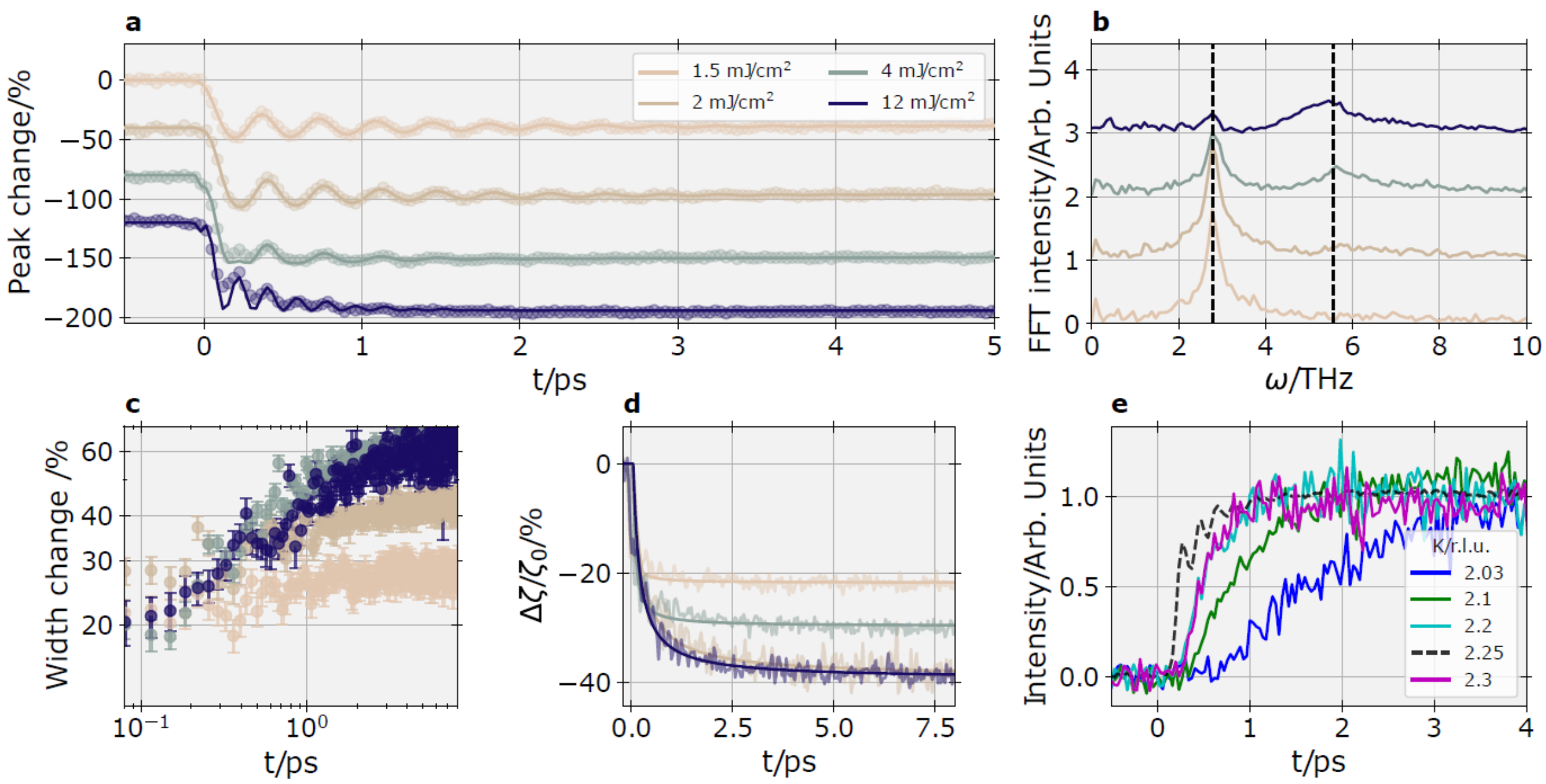

**Figure 3 Phonon dynamics during the melting of orbital order. a** Dynamics of the peak of the OTR at L=0.2, for various fluences, showing rapid and strong suppression and coherent oscillations. Solid lines are fits to the data using a spectator model (see **supplementary information**) **b** Fourier transform, FFT of the oscillations observed in **a** after subtracting the non-oscillatory background. At low fluences, the oscillations are observed at 2.8 THz. As the fluence increases, a second peak grows at the second harmonic of the fundamental. Dashed lines indicate the fundamental and its second harmonic. During melting, only a small shift in the phonon frequency and lifetime are found (see **extended data Figure 3**). **c** Dynamics of the OTR width as a function of fluence after subtracting the oscillatory signal on a log-log scale. The dynamics of the width increase is independent of fluence at early times, with the fluence affecting only the long-time values. Error bars are plus or minus one standard deviation obtained from the covariance of the fitting procedure. **d** Change in correlation length obtained from the reciprocal of the data in **c**. Solid lines are fits based on a rate equation model (see supplementary information). **e** Normalised dynamics of the diffuse scattering around the OTR (K=2.25, dashed line) at L=0.2 and 12 mJ/cm$^2$ excitation, showing a range of timescales. The K values are indicated by the markers of the same colour in **Fig.2f**.

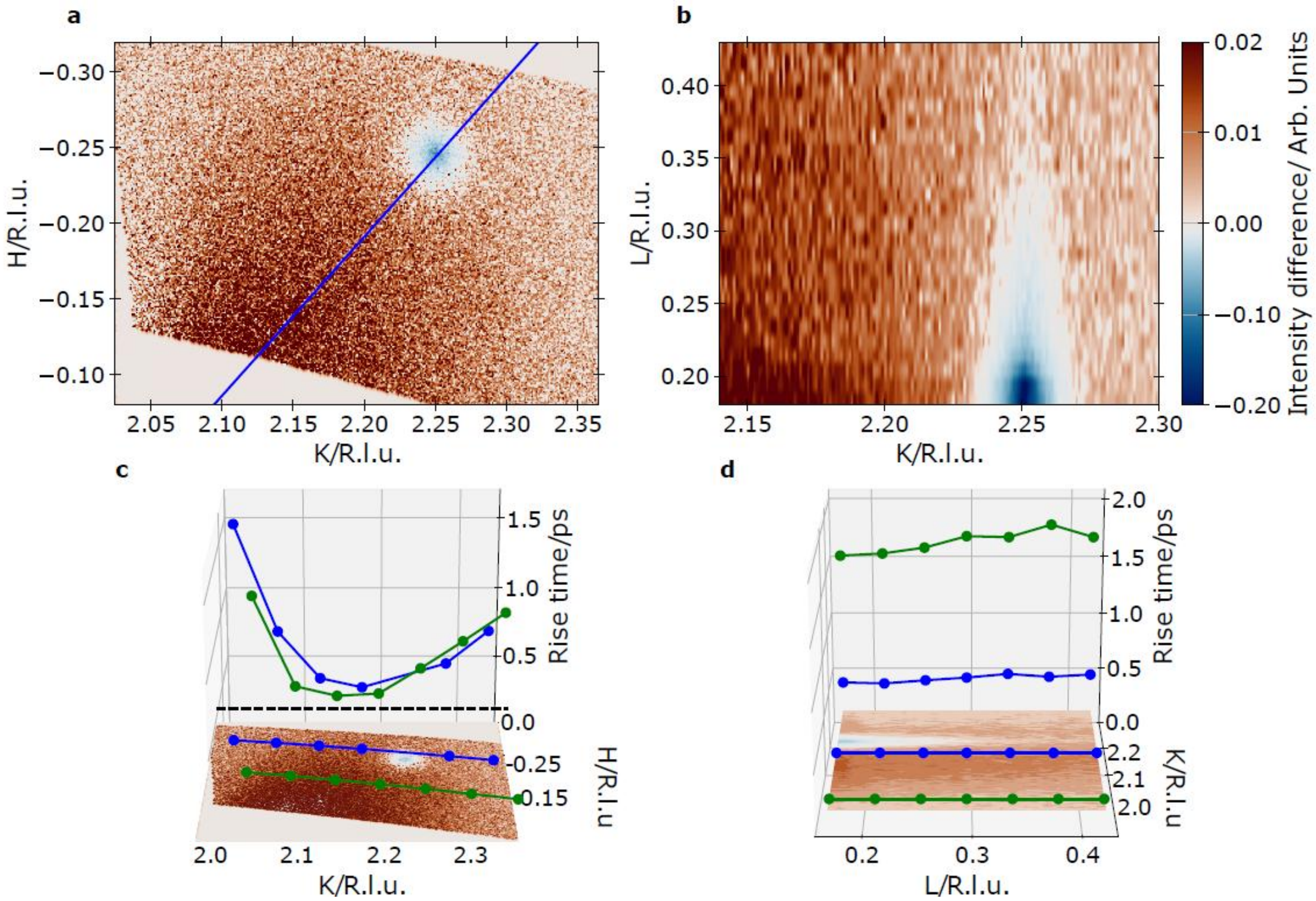

**Figure 4 Diffuse scattering during the melting of orbital order. a** Cuts in reciprocal space of the change in scattering around the OTR in the HK plane at L = 0.2, and **b** cut along the Line *[H, K]=[2-K, K]* (blue line in **a**) in the L direction. In addition to the suppression of the OTR (blue) a large region of diffuse scattering appears which increases in intensity as the (020) structural peak is approached. **c** Extracted timescales based on fitting the diffuse dynamics with $I(k,t) = A_k(1 - e^{-\frac{t}{\tau_k}})$ for delays greater than zero in the HK plane at L=0.2 **,** and **d** in the KL plane at H~-0.25. Dynamics are fastest along the line *[H, K]=[2-K, K]* and reduces further as the (020) structural peak is approached. Fluence dependent data is shown in **extended data Figure 5**.

## Extended Data

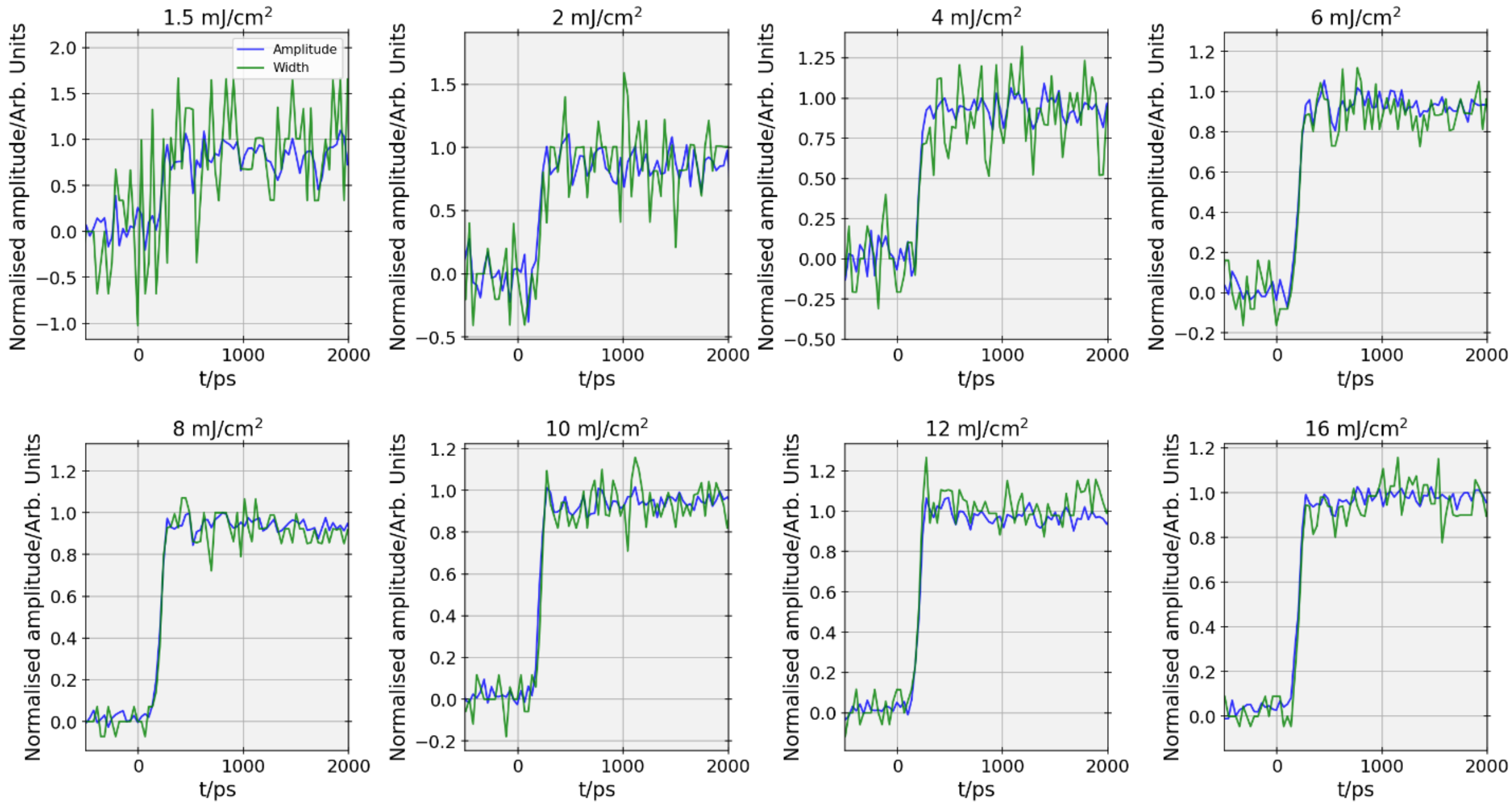

**Extended Data Figure 1** Amplitude and width dynamics of the orbital Bragg peak. For all fluences the peak response and the width follow the same normalized dynamics.

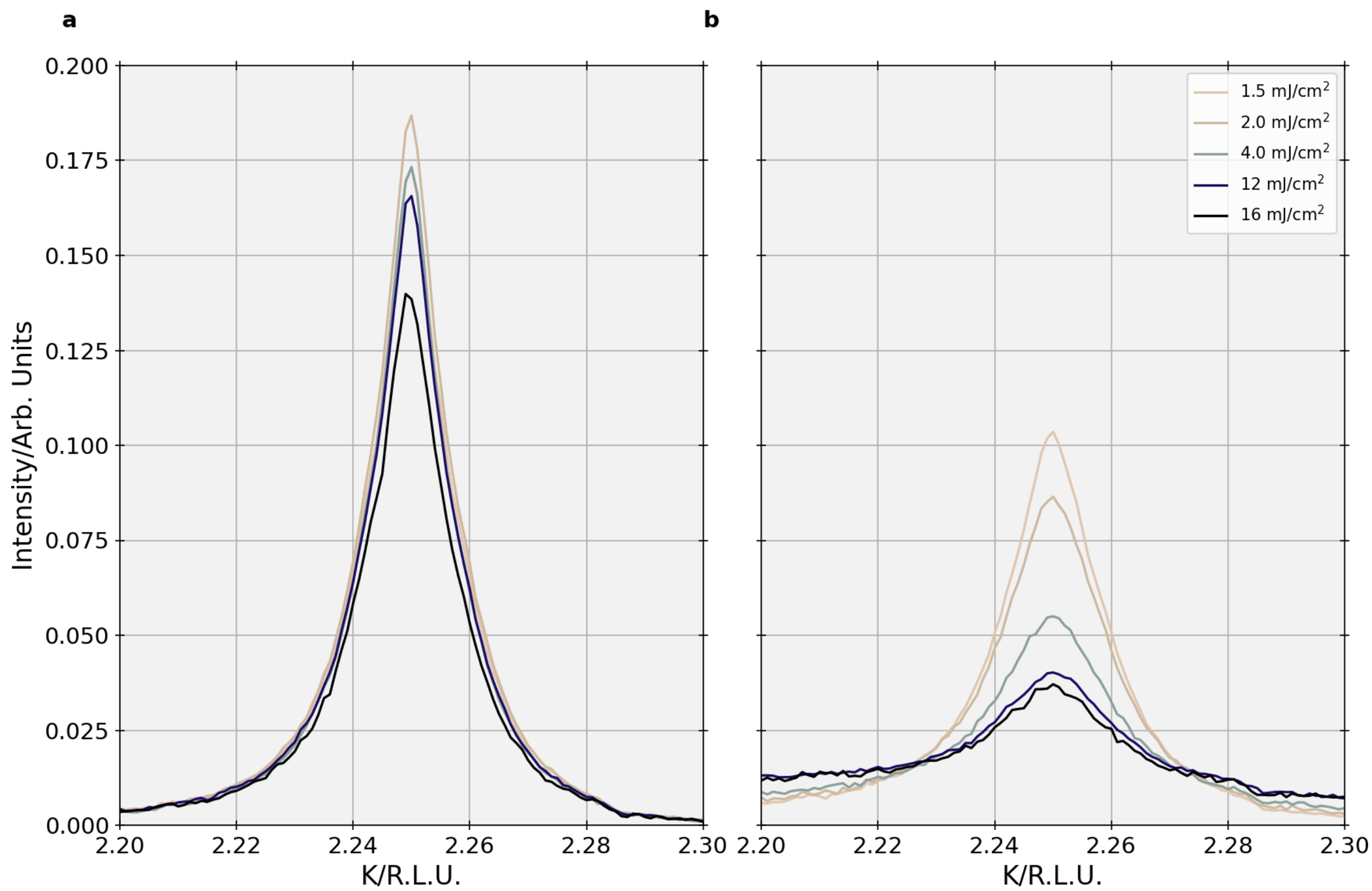

**Extended Data Figure 2** Permanent change in OTR surface order after 16 mJ/cm$^2$. X-ray measurements were made at 100 Hz, while the laser pumping was at 50 Hz. a The measured surface truncation rod at L=0.2 in the laser blocked state. The range 1.5-12 mJ/cm$^2$ show no significant change in intensity, indicating the sample fully recovers between pulses, with the small fluctuations due to FEL intensity and pointing fluctuations. However, for 16 mJ/cm$^2$ a significant reduction in peak height is seen which is not related to FEL fluctuations. b laser on channel (8.7 ps after excitation), the absolute pump induced effect at 16 mJ/cm$^2$ is not significantly lower than at 12 mJ/cm$^2$. Likely long range orbital order has been fully suppressed and only local distortions, which can persist above Tc are present. Note the OBP showed no noticeable effect at this fluence, indicating that the changes are occurring only at the surface.

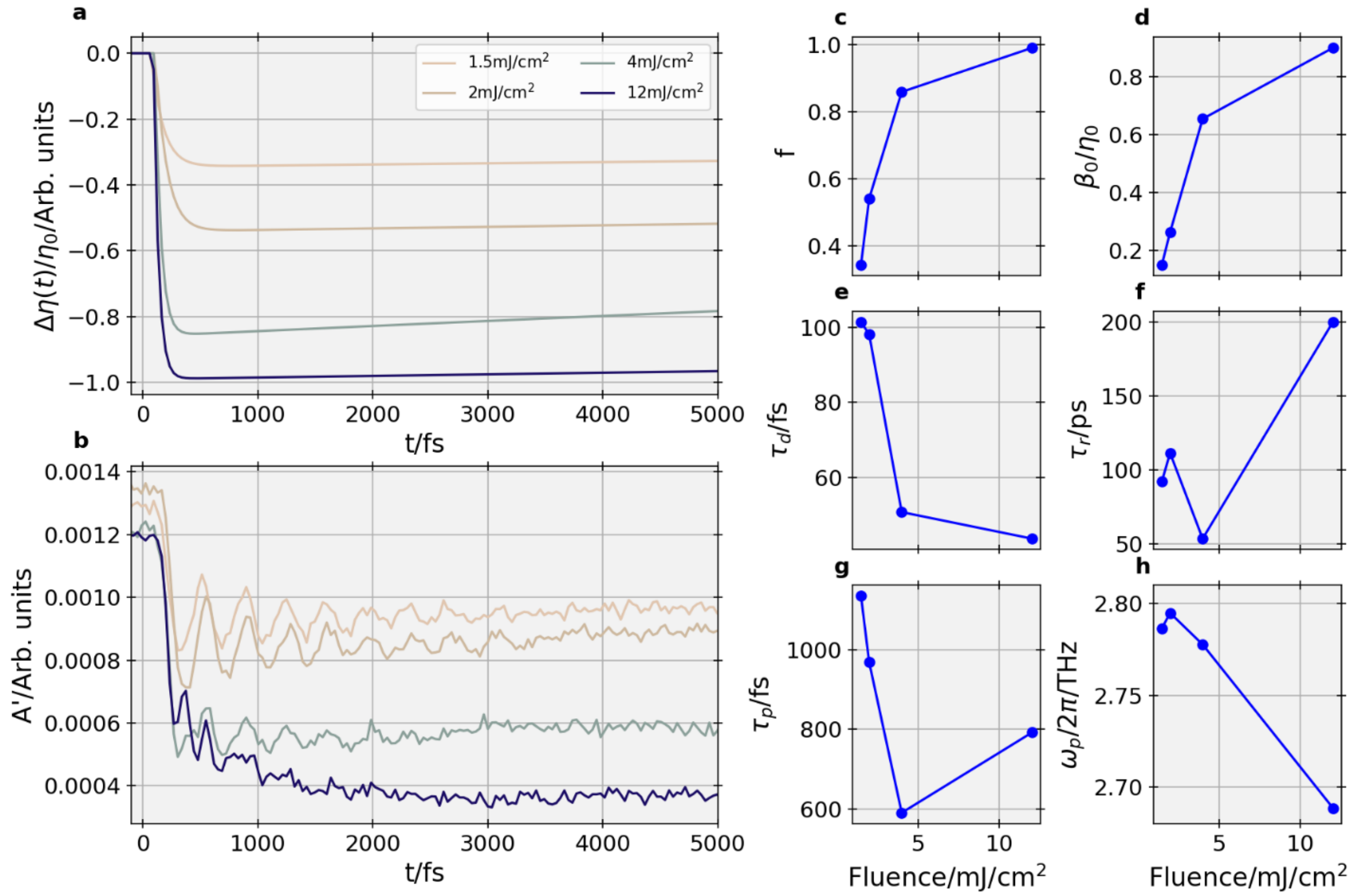

**Extended Data Figure 3** Extraction of order parameter from the orbital truncation rod. a Shows the extracted change in order parameter from the change in amplitude of the orbital truncation rod. b plots the rod area parameter, $A'$, which is proportional to the integrated area of the Lorentzian fitted to the transient data. c-h Fit parameters for the order parameter and phonon model. The parameters are defined in equation 1.

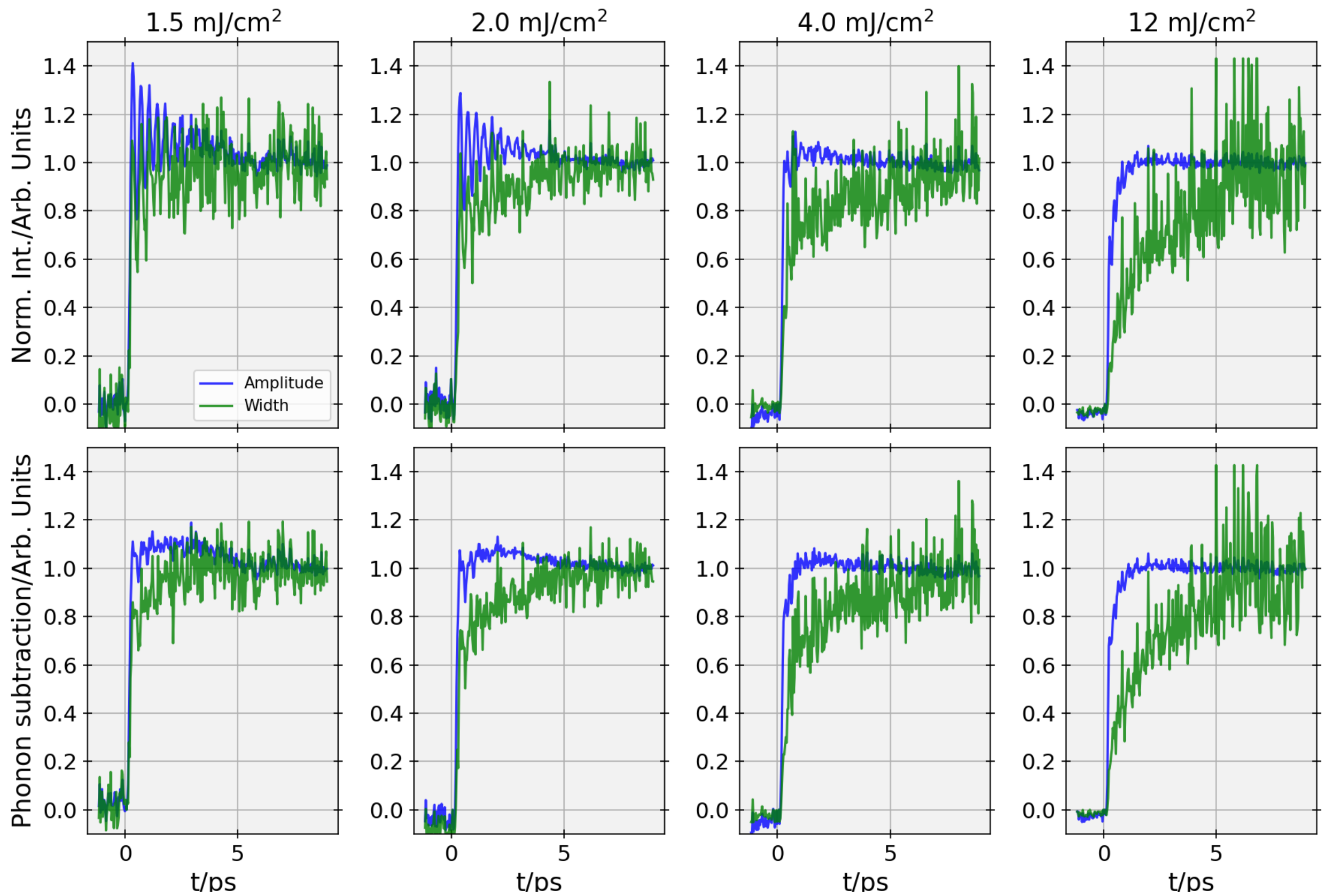

**Extended Data Figure 4** Comparison of the normalized amplitude and width dynamics of the OTR at L=0.2. The dynamics of the peak (blue) and width (orange) of the Lorentzian profile with (bottom row) and without (top row) subtraction of the spectator phonon mode.

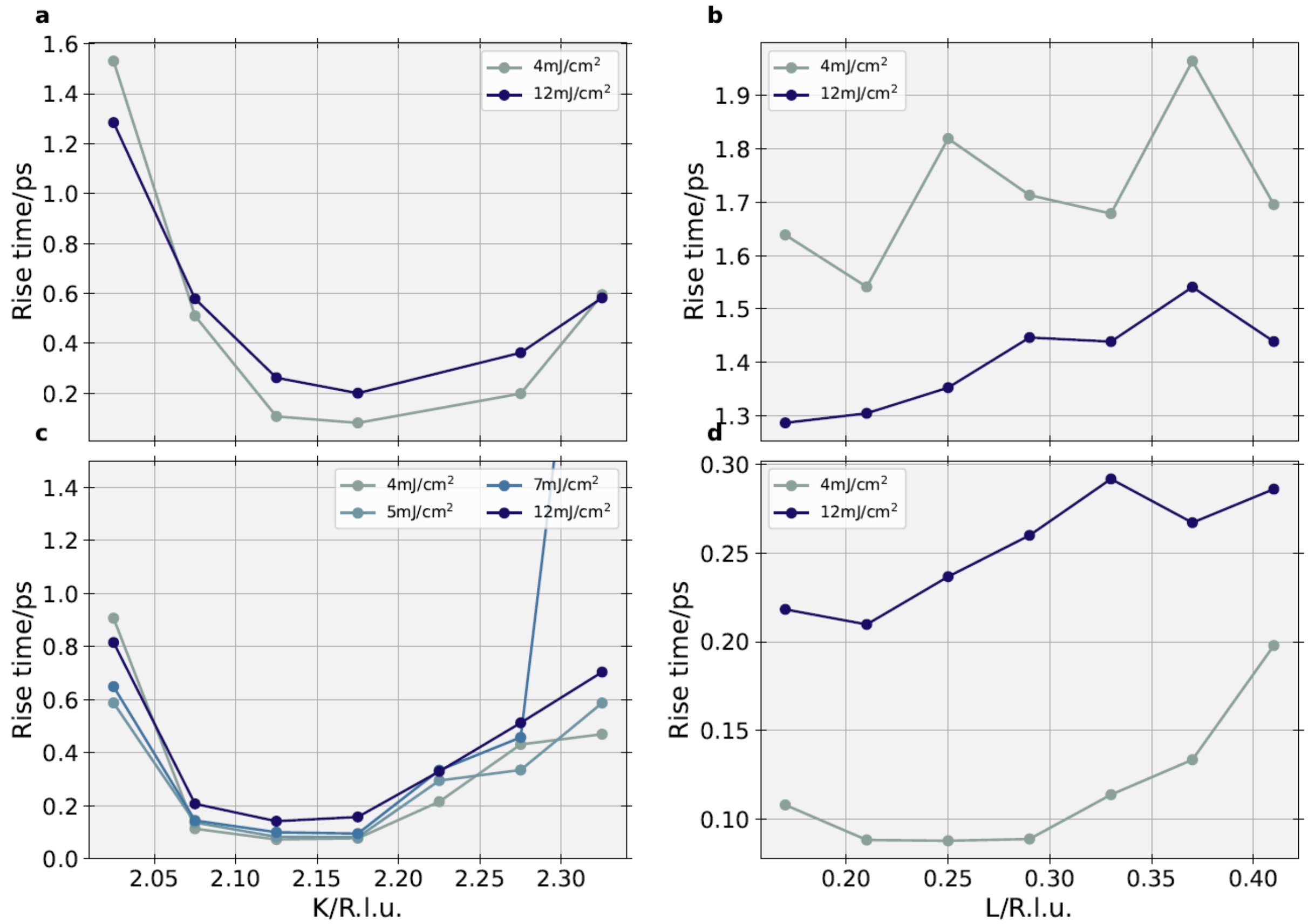

**Extended Data Figure 5** Comparison of the rise time of the diffuse scattering resulting from different fluence excitation. a/c show the fluence dependence of Fig 4c at L=0.2 (a – blue cut at H~0.25, b – green cut at H ~0.15). b/d shows the fluence dependence of Fig 4d at (b – greet cut at K=2, d – blue cut at K = 2.15).

## Methods

### Experimental setup

Experiments were performed at the SwissFEL Bernina instrument[33]. Single crystal samples were grown by the floating zone method and were cleaved in air to expose a (001) surface normal and mounted onto a 6-circle diffractometer. To achieve surface sensitivity, experiments were performed at a grazing incidence angle of 0.3° with a non-resonant photon energy of 12 keV, which was monochromatised to a 1.7 eV bandwidth. The base temperature of 180 K was used to avoid generation of any meta-stable states, which have been observed in lab-based optical experiments for temperatures below 140 K and to ensure the sample recovers between pump pulses. X-rays were focused to a spot of $2 \times 100\ \mu m^2$ onto the sample (normal incidence spot size).

The system was pumped with 800 nm laser pulses incident at 5° with a beam spot of $60 \times 2000\ \mu m^2$. The estimated time resolution of the setup was 70 fs (FWHM)[33]. The reported fluences are obtained from dividing the pulse energy by the tilted spot size and we do not correct for reflective losses.

In this geometry, the calculated X-ray penetration depth (110 nm) is nominally well matched to the optically pumped depth (116 nm) based on literature values[50,51] for the optical and X-ray constants. Despite this matching, this still means that the X-rays will probe regions of the sample that are only weakly excited.

We performed two types of measurements, time-scans at fixed detector settings, which captured cuts primarily in the KL plane of the orbital order and orbital truncation rod, and reciprocal space maps at fixed delays which scanned the sample around the surface normal, keeping the angle of incidence fixed.

Measurements of the OBP and OTR were performed at different $\phi$ angles (see Fig 1) and with the detector rotated to the appropriate angles. The incident angle did not change in this geometry.

**Pump and probe penetration depth**

The X-ray penetration depth was calculated using the data in Ref. [51]. The penetration of the pump was calculated use a refractive index of 1.36+0.37i, obtained from the reflectivity data of Ishikawa et al.[50] (180 K and for S-polarized, 800 nm beam). Taking into account refraction of the pump, the resulting in a penetration depth of 116 nm is obtained.

**Calculation of the in-plane correlation length and surface roughness from the orbital truncation rod**

The intensity of the orbital truncation rod can be expressed as[35]

$$I(K,L) = \left|\left(\frac{f(H,K,L)}{1-e^{2i\pi L-\alpha}}\right)\right|^2 \frac{A}{\sin\theta}\frac{L^2 r}{32(\Delta H^2 + \Delta K^2 + 4\pi^2 L^4 r^2)^{\frac{3}{2}}}$$

Where, $r = \frac{(2\sigma)^2}{C_r}$, $\alpha = \frac{\mu c}{2}\left(\frac{1}{\sin\alpha_i} + \frac{1}{L\frac{\lambda}{c} - \sin\alpha_i}\right)$, $\sigma$ is the surface roughness, $c_r$ is the phase-slip length. $\alpha_i = 0.3°$ is the incident angle, μ is the absorption coefficient, c is the lattice constant along the out of plane direction and λ is the X-ray wavelength. $f(H,K,L)$ is the scattering factor of the orbital ordering supercell.

After averaging the orbital ordering domain configurations for $f(H,K,L)$, the above equation can be approximated by a Lorentzian function of K in the KL plane:

$$I(K,L) = \left|\left(\frac{1}{1-e^{2i\pi L-\alpha}}\right)\right|^2 A\frac{L^2 r}{8(K-K_0)^2 + 2\pi^2 L^4 r^2}$$

We performed a reciprocal space map of the orbital order to build up a three-dimensional dataset to get the intensity as a function of H,K,L. We then integrate along the H direction to get a two-dimensional representation. The dataset is fitted with two parameters for all L values: A, representing the amplitude of the distortion and $r$=0.039, after a background has been removed.

The shape of the rod at a fixed L can be described with a Lorentzian line shape:

$$I(K, L = 0.2) = \frac{\Gamma A'}{(K - K_0)^2 + \frac{\Gamma^2}{4}}$$

From which we can extract an in-plane correlation length:

$$\zeta_s = {}^{a}/_{2\pi\Gamma} \sim 40\text{Å}$$

Where $a = 3.853\text{Å}$ is the in-plane lattice constant.

**Fitting the Orbital Bragg peak**

To determine the correlation lengths of the bulk orbital order from the OBP, we carried out a reciprocal space scan as mentioned in the main text. To extract the in-plane correlation length, we selected the (H K 2) map and fit the data in H and K directions using a Lorentzian profile as above. To this end, the data were first converted from reciprocal space unit (*r.l.u)* into *q*-space and then fitted. The correlation length of the OBP, is then the reciprocal of the extracted width (FWHM). This yielded a value of 250 Å for the in-plane correlation length.

For the out-of-plane correlation length, each image from the RSM in the HK plane was binned in H or K direction and maximum of the resulting peak was taken as a function of L and then fit with the same functional.

For the time-resolved scans, on the other hand, the CCD was fixed and we, therefore, took a cut through the OBP along (-2.25 K 2) and (H 2.25 2), respectively, and fit the traces at each time delay.

## Supplementary Information

**Spectator modes**

In charge density wave compounds, oscillations are often observed in the dynamics of the Bragg peaks. At low fluences the peak oscillates around a distortion of finite amplitude. However, at high fluence, the sudden appearance of frequency upshift is attributed to the order parameter oscillating around a high symmetry state with no average distortion[14–16]. As the peak intensity is proportional to the square of the distortion, or order parameter, the peak oscillates at twice the frequency of the actual motion when oscillations occur in the high symmetry state. However, we do not believe this to be the case here. Firstly, before the frequency increases, the order parameter frequency should soften, which is not seen. Secondly, it would be highly coincidental if the high symmetry potential had the same curvature as the symmetry broken state. Finally, the oscillation does not strongly modulate the OBP measured here, which should also be sensitive to the order parameter dynamics.

Instead, we suggest that the observed oscillation is due to a spectator phonon. The 2.8 THz mode seen here can be assigned to the motion of the La/Sr ions[37], which do not appreciably change position during the phase transition and are thus not directly related to the order parameter[52]. The reduction in symmetry from *I4/mmm* at high temperatures to *Cmcm* in the orbitally ordered state causes a large increase in the number of Raman active modes. Phonons from the M or LD points of the high symmetry phase get mapped onto the $\Gamma$-point, increasing the number of modes from Raman active modes from six to 39. These new 'back-folded' modes can be considered spectator modes[11,22] of the phase transition. As the manganites have strong electron-phonon coupling, ultrafast electronic excitation can be expected to drive large amplitude coherent oscillations of these new modes[22,53] and X-ray scattering from these back-folded modes will then modulate the intensity at the same wavevector as the orbital ordering.

Unlike in equilibrium, where the scattering at the orbital ordering wavevector is only proportional to the squared magnitude of the order parameter, the out-of-equilibrium scattering will now contain an additional contribution from the coherent phonon. If the order parameter is not suppressed, X-rays scattered from the orbital order and the phonon will interfere, resulting in a term contributing to the scattering which is linear in the phonon displacement. However, if the orbital order is fully suppressed during the dynamics, the interference is lost and the mode appears at the second harmonic, as is the case for phonon modes at a finite wavevector[54].

To describe the change in the phonon frequency we present a simple toy model for the "spectator mode", which is not meant to capture the true structural changes, but highlights how the phonon frequency shifts to the second harmonic. We take simplified unit cell consisting of only La/Sr ions located at (0,0,0) and the Mn ions at (½, ½, 0). We assume that the order parameter can be represented by a displacement of the Mn ions by an amount $\eta$. In equilibrium, this will generate orbital order peaks whose intensity, $I_{OO}$, will scale as $I_{OO} \propto \eta^2$ for small displacements.

If the La/Sr ions remain at the high symmetry position, they do not contribute to any scattering to the $I_{OO}$ peaks. However, the displacement of the Mn ions modifies the environment of the La/Sr ions, so that these atoms can also become Raman active, and can thus move away from their high symmetry points. If the La/Sr ions oscillate by an amount $\beta$ at the same wavevector as the order parameter, the orbital order intensity will then be given by the coherent sum of the two scattering factors or

$$I_{OO} \propto (\eta + \beta)^2$$

We assume that the La/Sr ions are initially in the high symmetry positions and coherent oscillations are generated with a sine-like phase such that $\beta(t) = \beta_0 e^{-t/t_p} \sin \omega_p t$, i.e. they oscillate around their equilibrium position, rather than a non-equilibrium position. We then assume that the order parameter follows a non-oscillatory dynamic, $\eta(t)$. The change in superlattice intensity can then be written as

$$\begin{aligned}\Delta I_{SL} &= \eta^2(t) + \beta^2(t) + 2\eta(t)\beta(t) - \eta_0^2 \\ &= \eta^2(t) - \eta_0^2 + \frac{\beta_0^2}{2} + 2\eta(t)\beta_0 e^{-t/t_p} \sin \omega_p t - \frac{\beta_0^2}{2} e^{-t/2t_p} \cos 2\omega_p t\end{aligned}$$

Here it can be seen that if $\eta(t)$ remains finite, the orbital order intensity will oscillate at $\omega$, whereas if the order parameter is suppressed $\eta(t) \to 0$ the frequency changes the $2\omega$.

To fit the dynamics of the intensity, we assume that in addition to the scattering from the order parameter $\eta$ and the phonon $\beta$ there is some incoherent background, $B$, which can be noise on the detector or other scattering which is independent of any pump parameter. This gives the measured intensity as

$$I_{OO} \propto (\eta + \beta)^2 + B$$

The intensity before pumping is given by

$$I_{OO}(-t) \propto \eta_0^2 + B$$

Where $\eta_0$ is the order parameter value before pumping.

The measured relative change is then given as

$$\frac{\Delta I_{OO}}{I_{OO}(-t)} = \frac{(\eta+\beta)^2 - \eta_0^2}{\eta_0^2 + B}$$

When then assume that $\eta(t)$ has the same functional form as extracted from optical measurements[22]

$$\eta(t) = \eta_0 \left(1 - \frac{\delta\eta}{\eta_0}(t)\right)$$

Where

$$\frac{\delta\eta}{\eta_0}(t) = f\left(1 - e^{-t/\tau_d}\right)e^{-t/\tau_r}$$

and $\tau_d$ and $\tau_r$ are the decay and recovery time of the order and $f$ is the melt fraction. Substituting this into the equation for the relative change in superlattice intensity gives

$$\frac{\Delta I_{SL}}{I_{SL}(-t)} = \frac{\eta_0^2}{\eta_0^2 + B}\left[-\left(2 - \frac{\delta\eta}{\eta_0}\right)\frac{\delta\eta}{\eta_0} + \frac{\beta_0^2}{2\eta_0^2} e^{-\frac{2t}{\tau_p}}\left(1 - \cos 2\omega_p t\right) + \frac{2\beta_0}{\eta_0}\left(1 - \frac{\delta\eta}{\eta_0}\right) e^{-t/\tau_p} \sin \omega_p t\right]. \quad (1)$$

This model is then fitted to the peak intensity shown in Fig 2c. The corresponding dynamics for the order parameter are plotted in extended data figure 2. Note that the parameter B is the same for all fluences.

We note that unlike Ref 13, we do not need to include a time-dependent dephasing constant to fit our data because the order parameter dynamics and the oscillations are not coupled. We find that the phonon displacement scales roughly linear with applied fluence before saturating. This points to the role played by the order parameter in enabling the excitation of the mode. Once the order parameter is fully suppressed, the phonon mode can no longer be excited in the linear regime.

Although the phonon is only observed on the OTR, it is not a surface phonon as it is also seen in optical reflectivity data[22] and is seen in bulk Bragg peaks in other materials[11,17,18]. The lack of phonon in the OBP measured here, indicates the phonon has a dispersion along L, i.e. bulk peaks at different wavevectors are likely to show the phonon.

Dynamics of the in-plane correlation length

As the OTR signal is convolved with the phonon response, we need to remove the phonon signal at all momenta. While this could be achieved with the equations shown in the main text, it would require assumptions about the K dependence of the phonon and the order parameter. Instead, we adopt a phenomenological approach. We first perform a PCA analysis on the OTR data and find that the phonon appears mainly in the first 3 components. We then fit the phonon as a summation of sine waves and subtract this from the raw data. We then fit the phonon subtracted raw data to extract the width shown in Fig 3c. A comparison of the width dynamics with and without phonon subtraction are shown in extended data figure 4.

We convert the fitted width to a correlation length via $\zeta = a/2\pi\Gamma$. The general solution to the equation for the correlation length is then

$$\zeta(t) = \zeta_p + \frac{\zeta_0 - \zeta_p}{1 + k(\zeta_0 - \zeta_p)t}$$

Where $\zeta_0$ is the initial correlation length. Fig. 3d plots the change in correlation length $\frac{\Delta\zeta(t)}{\zeta_0} = \frac{\zeta(t)}{\zeta_0} - 1$, which can be written using the above formula and defining $\Delta\zeta_p = \zeta_0 - \zeta_p$ as

$$\frac{\Delta\zeta(t)}{\zeta_0} = -\frac{\Delta\zeta_p}{\zeta_0}\left(1 - \frac{1}{1 + k\Delta\zeta_p t}\right)$$

## Data availability

The relevant data supporting the findings of this study are available via Zenodo at 10.5281/zenodo.15708245 (ref. 55)

## Method-only references